\documentclass[11pt]{article}
\usepackage{amsmath}

\usepackage{longtable}
\usepackage[english]{babel}

\begin{document}
\title{\Large\bf Assignment of Framework Types to the Zeolite Crystals in the Inorganic Crystal Structure Database}
\author{M. Lach-hab, S. Yang, I. Vaisman, and E. Blaisten-Barojas \\
Computational Materials Science Center, \\George Mason University, MSN
6A2, \\Fairfax, Virginia 22030}
\maketitle

\begin{abstract}
In this work the framework type was assigned to 1370 zeolite crystals included in the Inorganic Crystal Structure Database.
\end{abstract}
\section{Table of Framework Types of Zeolite Crystals}

The Inorganic Crystal Structure Database (ICSD) \cite{ICSD} has the most complete crystal structure collection gathered up to date. Table I contains 1370 zeolite crystals from the ICSD. The ICSD provides bibliographic and crystallographic information for each entry, but contains no information about the framework type of each zeolite. In this work we assign the framework type codes (FTC)  to the 1370 zeolite database entries. The newly assigned codes are gathered in Table 1, which lists also the name of each crystal, its chemical formula and database entry number. The zeoTsites package \cite{Sastre} was used for determining the coordination sequences (CS) \cite{Meier} and vertex symbols (VS) \cite{MO} for 1370 zeolite. With the combined CS-VS information, 94 FTCs were assigned to these crystals according to rules set by the Structure Commission of International Zeolite Association \cite{IZA}. New findings in this table have been communicated to the National Institute of Standards and Technology for inclusion in the ICSD.

\begin{tiny}
\begin{center}
\begin{longtable}{|p{1.2cm}|p{5cm}|p{7cm}|p{1cm}|}

\caption{Framework Type Codes of Zeolite Crystals in the ICSD} \label{grid_mlmmh} \\

\hline 
\multicolumn{1}{|p{1.2cm}|}{\textbf{Framework Type Code}} & \multicolumn{1}{p{5cm}|}{\textbf{Chemical Name}} & \multicolumn{1}{p{7cm}|}{\textbf{Chemical Formula}} & \multicolumn{1}{p{1cm}|}{\textbf{ICSD Code}} \\ \hline
\endfirsthead

\multicolumn{4}{l}%
{{\bfseries \tablename\ \thetable{} -- continued from previous page}} \\
\hline \multicolumn{1}{p{1.2cm}|}{\textbf{Framework Type Code}} &
\multicolumn{1}{p{5cm}|}{\textbf{Chemical Name}} &
\multicolumn{1}{p{7cm}|}{\textbf{Chemical Formula}} &
\multicolumn{1}{|p{1cm}|}{\textbf{ICSD Code}} \\ \hline
\endhead

\hline \multicolumn{4}{|l|}{{Continued on next page}} \\ \hline
\endfoot

\hline \hline
\endlastfoot

ABW & Lithium Tecto-alumosilicate Hydrate & Li0.91(Al0.91Si1.09O4) (H2O)0.98 & 6269  \\
ABW & Thallium Tecto-alumosilicate (1.1/1) & Tl1.1(AlSiO4) & 66153  \\
ABW & Lithium Tecto-alumogermanate Hydrate & Li(AlGeO4) (H2O) & 91662  \\
ABW & Ammonium Zincoarsenate(V) & (NH4)(ZnAsO4) & 93115  \\
ABW & Lithium Tecto-alumosilicate & Li(AlSiO4) & 97909  \\
ABW & Lithium Tecto-alumosilicate & Li(AlSiO4) & 97910  \\
ABW & Lithium Tecto-alumosilicate Dideuteriooxide & Li(AlSiO4)(D2O) & 60892  \\
ABW & Lithium Tecto-alumosilicate Hydrate & Li(AlSiO4) (H2O) & 68257  \\
\end{longtable}
\end{center}
\end{tiny}

\begin{center}
\bf {Acknowledgments}
\end{center}

\noindent This work was supported by the National Science Foundation grant CHE-0626111. Authors acknowledge the NIST Standard Reference Data Program for making available the ICSD zeolite data set. EBB acknowledges useful discussions with Dr. V. Karen of the National Institute of Standards and Technology.

\end{document}